\documentclass[reprint,aps,prl,superscriptaddress]{revtex4-2}

\usepackage[pdftex]{graphicx}
\usepackage{amsmath}

\begin{document}

\title{Electronic nematic order in the normal state of strontium ruthenate}

\author{Ryan Russell}
\affiliation{Materials Department, University of California, Santa Barbara, California 93106, USA}

\author{Hari P. Nair}
\affiliation{Department of Materials Science and Engineering, Cornell University, Ithaca, New York 14853, USA}

\author{Kyle M. Shen}
\affiliation{Physics Department, Cornell University, Ithaca, New York 14853, USA}
\affiliation{Kavli Institute at Cornell for Nanoscale Science, Ithaca, New York 14853, USA}

\author{Darrell G. Schlom}
\affiliation{Department of Materials Science and Engineering, Cornell University, Ithaca, New York 14853, USA}
\affiliation{Kavli Institute at Cornell for Nanoscale Science, Ithaca, New York 14853, USA}
\affiliation{Leibniz-Institut f\"{u}r Kristallz\"{u}chtung, 12489 Berlin, Germany}

\author{John W. Harter}
\email[Corresponding author: ]{harter@ucsb.edu}
\affiliation{Materials Department, University of California, Santa Barbara, California 93106, USA}

\date{\today}

\begin{abstract}
Despite significant achievements in characterizing the properties of Sr$_2$RuO$_4$ over the last three decades, the precise nature of its electronic ground state is still unresolved. In this work, we provide a missing piece of the puzzle by uncovering evidence of electronic nematic order in the normal state of Sr$_2$RuO$_4$, revealed by ultrafast time-resolved optical dichroism measurements of uniaxially strained thin films. This nematic order, whose domains are aligned by the strain, spontaneously breaks the four-fold rotational symmetry of the crystal. The temperature dependence of the dichroism resembles an Ising-like order parameter, and optical pumping induces a coherent oscillation of its amplitude mode. A microscopic model of intra-unit-cell nematic order is presented, highlighting the importance of Coulomb repulsion between neighboring oxygen $p$-orbitals. The existence of electronic nematic order in the normal state of Sr$_2$RuO$_4$ may have consequences for the form and mechanism of superconductivity in this material.
\end{abstract}

\maketitle

The study of Sr$_2$RuO$_4$ entered a new era after the recent observation of a reduction in the Knight shift at the onset of superconductivity~\cite{pustogow2019}, which overturned the long-favored chiral $p$-wave triplet pairing scenario. Researchers are presently faced with an array of sometimes contradictory results, and it is clear that a fundamental piece of the Sr$_2$RuO$_4$ puzzle is still missing~\cite{mackenzie2017}. A complete understanding of the superconducting phase in any material depends upon a detailed knowledge of the normal state electronic structure out of which it emerges. It has long been believed that the normal state of Sr$_2$RuO$_4$ is a conventional quasi-two-dimensional Fermi liquid with moderate correlations~\cite{mackenzie2017,mackenzie2003}. Evidence of any deviation from conventional Fermi liquid behavior would have significant consequences for both theoretical proposals and the interpretation of experiments. In this work, we present such evidence. Motivated by recent transverse resistivity experiments indicating possible electronic nematicity~\cite{wu2020}, we use optical dichroism measurements of high-quality epitaxially strained thin films to uncover evidence of electronic nematic order in the normal state of Sr$_2$RuO$_4$.

\begin{figure*}[t]
\includegraphics{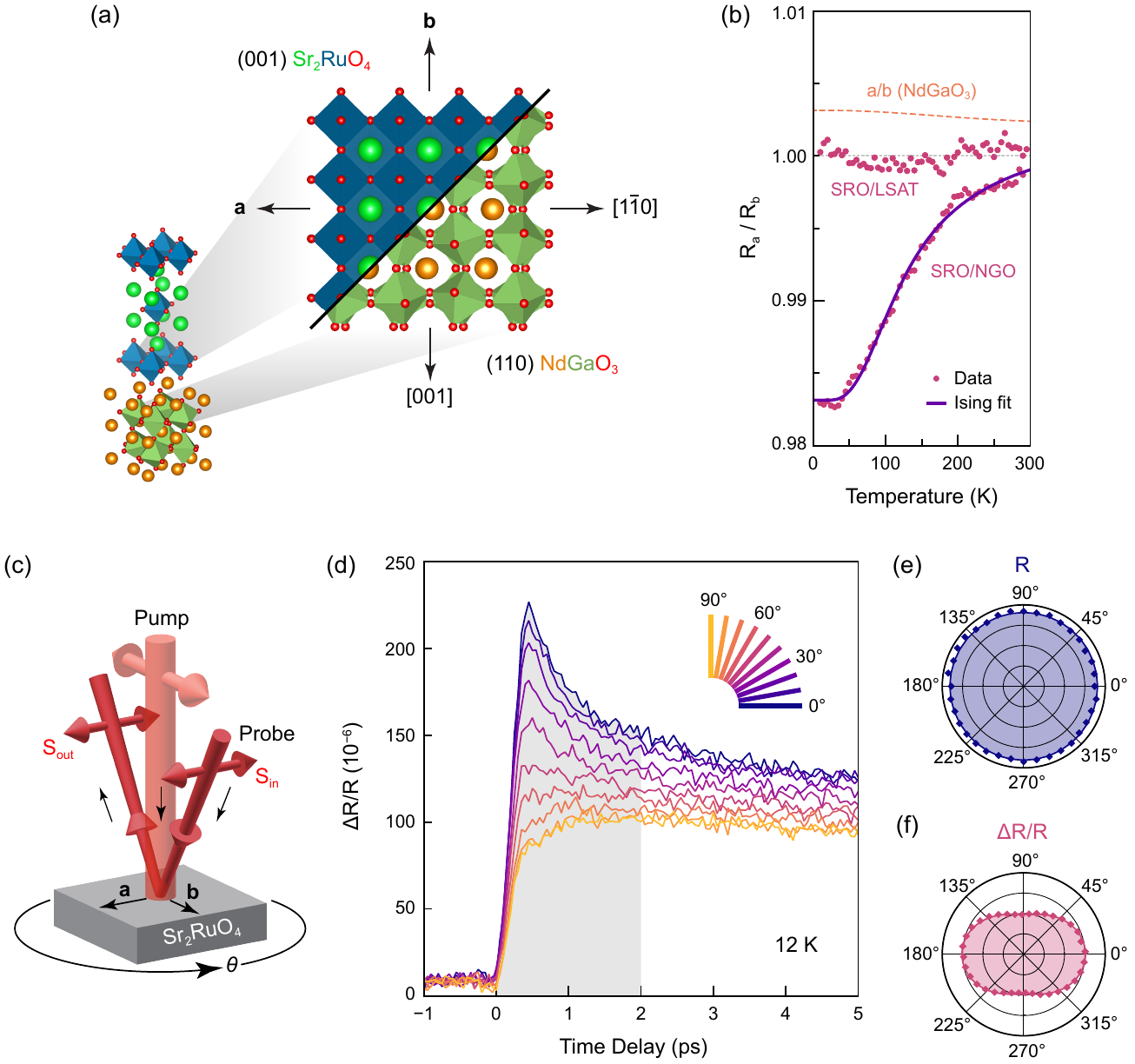}
\caption{\label{Fig1} Static and transient optical measurements of Sr$_2$RuO$_4$. (a)~Crystallographic directions in the film and substrate. For Sr$_2$RuO$_4$ on (110) NdGaO$_3$, the $a$-axis corresponds to [1$\bar{1}$0] NdGaO$_3$ and the $b$-axis corresponds to [001] NdGaO$_3$. (b)~Static reflectivity anisotropy $R_a/R_b$ versus temperature for films grown on NdGaO$_3$ (SRO/NGO) and LSAT (SRO/LSAT). A fit to an Ising model of nematic order is shown for the NdGaO$_3$ sample. Also shown is the temperature-dependent lattice constant ratio $a/b$ for films strained to NdGaO$_3$ from Ref.~\citenum{savytskii2003}. (c)~Illustration of the pump-probe transient reflectivity experimental setup. The pump and probe beams remain fixed in a cross-polarized configuration while the sample is rotated about its surface normal. $\theta = 0^\circ$ when the probe is polarized along the $a$-axis (as shown) and $\theta = 90^\circ$ when polarized along the $b$-axis. (d)~Raw transient reflectivity curves for several values of $\theta$ between $0^\circ$ and $90^\circ$ at 12~K. A clear anisotropy is observed in the picosecond-scale transient optical response, with a peak at $0^\circ$ and a flat step at $90^\circ$. (e)~Polar plot of static reflectivity versus $\theta$, showing approximate isotropy. (f)~Polar plot of $\Delta R/R$ integrated from 0 to 2~ps [gray shaded region in (c)] versus $\theta$, showing pronounced anisotropy.}
\end{figure*}

Nematic order spontaneously breaks rotational symmetry while preserving translational symmetry. Nematicity in unconventional superconductors is far from rare, occurring in both the iron-based superconductors and the cuprates~\cite{chuang2010,daou2010,bohmer2022}. A nematic transition in these systems, however, is usually accompanied by a structural or magnetic transition. Such a transition is absent in Sr$_2$RuO$_4$. In a tetragonal crystal, nematicity results in a lowering of symmetry to an orthorhombic unit cell where the $a$ and $b$ directions become nonequivalent. The spontaneous nature of the symmetry breaking typically leads to the formation of opposing domains related by 90$^\circ$ rotations, whose average over long length and time scales retains the parent tetragonal symmetry. Thus, directly detecting electronic nematic symmetry breaking with macroscopic probes generally requires stabilizing the formation of one type of domain over the other, akin to structural detwinning. Uniaxial strain is an ideal external conjugate field by which one can achieve this effect. In the iron-based superconductors, for example, uniaxial strain has been used successfully to study nematic order with optical, transport, and photoemission experiments~\cite{dusza2011,tanatar2010,chu2010,yi2011,chinotti2017}. Here, we take advantage of uniaxial epitaxial strain to enable macroscopic transient optical reflectivity measurements of electronic nematic order in Sr$_2$RuO$_4$.

We study high-quality thin films of (001)-oriented Sr$_2$RuO$_4$ grown by molecular-beam epitaxy, as described in Ref.~\citenum{nair2018}. The films are coherently strained by growing them on (110) NdGaO$_3$ single-crystal substrates~\cite{savytskii2003}, guaranteeing a clean, uniform strain field over the entire sample area. Electrical transport measurements confirm the films are superconducting, with a critical temperature of $T_c \approx 1.4$~K (see Supplemental Materials~\cite{SM}). NdGaO$_3$ induces a small uniaxial ($B_{1g}$) strain of $a/b - 1 \approx 0.3\%$ at low temperatures, where $a$ and $b$ are the in-plane lattice constants of the coherently strained Sr$_2$RuO$_4$ film. Specifically, $a$ and $b$ correspond to the [1$\bar{1}$0] and [001] spacings, respectively, of the (110) NdGaO$_3$ substrate in the $Pbnm$ setting~\cite{savytskii2003}, as illustrated in Fig.~\ref{Fig1}(a). At room temperature, the uniaxial strain is reduced to $\sim$0.2$\%$. Strain has been used extensively in recent years to study the Lifshitz transition in Sr$_2$RuO$_4$ and to enhance the superconducting critical temperature~\cite{hicks2014,burganov2016,steppke2017,barber2019,sunk2019,grinenko2021,li2021}. We emphasize, however, that the magnitude of uniaxial strain that we employ is not sufficient to induce a Lifshitz transition (estimated in Ref.~\citenum{barber2019} to be $\varepsilon_\mathrm{xx} - \varepsilon_\mathrm{yy} \approx 0.7$\%), and moreover the epitaxial strain is biaxially compressive ($-0.1$\% in the $a$ direction and $-0.4$\% in the $b$ direction at low temperatures) relative to bulk Sr$_2$RuO$_4$~\cite{vogt1995,chmaissem1998}, which is known to move the Fermi level \textit{away} from the van Hove singularity rather than towards it~\cite{burganov2016}. Thus, the effects that we measure are due to a relatively weak uniaxial perturbation of the electronic structure and are not expected to be driven by density of states enhancements or changes in band topology.

To determine whether signatures of electronic nematic order are evident in the optical response of Sr$_2$RuO$_4$ at our probe wavelength of 800~nm, we first examine the static reflectivity anisotropy, which we define as the ratio of optical reflectivity with electric field polarized along the $a$-axis to that along the $b$-axis of the film. Figure~\ref{Fig1}(b) shows the measured reflectivity anisotropy as a function of temperature, where a striking dichroism is apparent. We find a decrease in reflectivity along the (long) $a$-axis relative to the (short) $b$-axis, which becomes more pronounced at lower temperatures. At the lowest temperatures, we measure a maximum reflectivity anisotropy of $1.7\%$, more than five times larger than the uniaxial strain imposed by the substrate ($0.3\%$). Similar behavior has been observed in detwinned FeSe, where the optical reflectivity anisotropy was taken as a proxy for the nematic order parameter~\cite{chinotti2017}. To eliminate the possibility that the dichroism is merely due to structural orthorhombicity, we include in the figure the temperature-dependent lattice constant ratio $a/b$ for films strained to NdGaO$_3$. Under the hypothesis that the reflectivity anisotropy is entirely due to simple $a \ne b$ structural effects, we would expect a linear proportionality with the substrate orthorhombicity as it changes with temperature (see Supplemental Materials~\cite{SM}). This is not what we observe. Instead, the dichroism grows by an order of magnitude upon cooling from room temperature to low temperature, while the substrate orthorhombicity only increases by $30\%$. As a control, we also measure the reflectivity anisotropy of a Sr$_2$RuO$_4$ thin film grown on (001) (LaAlO$_3$)$_{0.29}$(SrTa$_{1/2}$Al$_{1/2}$O$_3$)$_{0.71}$ (LSAT), which is tetragonal and induces no uniaxial strain ($a = b$)~\cite{steins1997}. In this case, we detect no appreciable dichroism. Our static optical anisotropy measurements suggest that the electronic structure of Sr$_2$RuO$_4$ possesses either spontaneous nematic order (scenario~I) or a large nematic susceptibility (scenario~II). In scenario~I, an electronic instability results in the formation of microscopic nematic domains fluctuating in space and time, with one orientation favored by the uniaxial strain over the other. This imbalance induces a net macroscopic anisotropy in the optical response of the electrons. In scenario~II, the weak uniaxial strain drives a strong electronic nematic response via a large nematic susceptibility. Without the uniaxial strain, however, the electronic structure would retain tetragonal symmetry even at low temperatures. Both scenarios are consistent with an absence of dichroism in films grown on LSAT.

\begin{figure*}[t]
\includegraphics{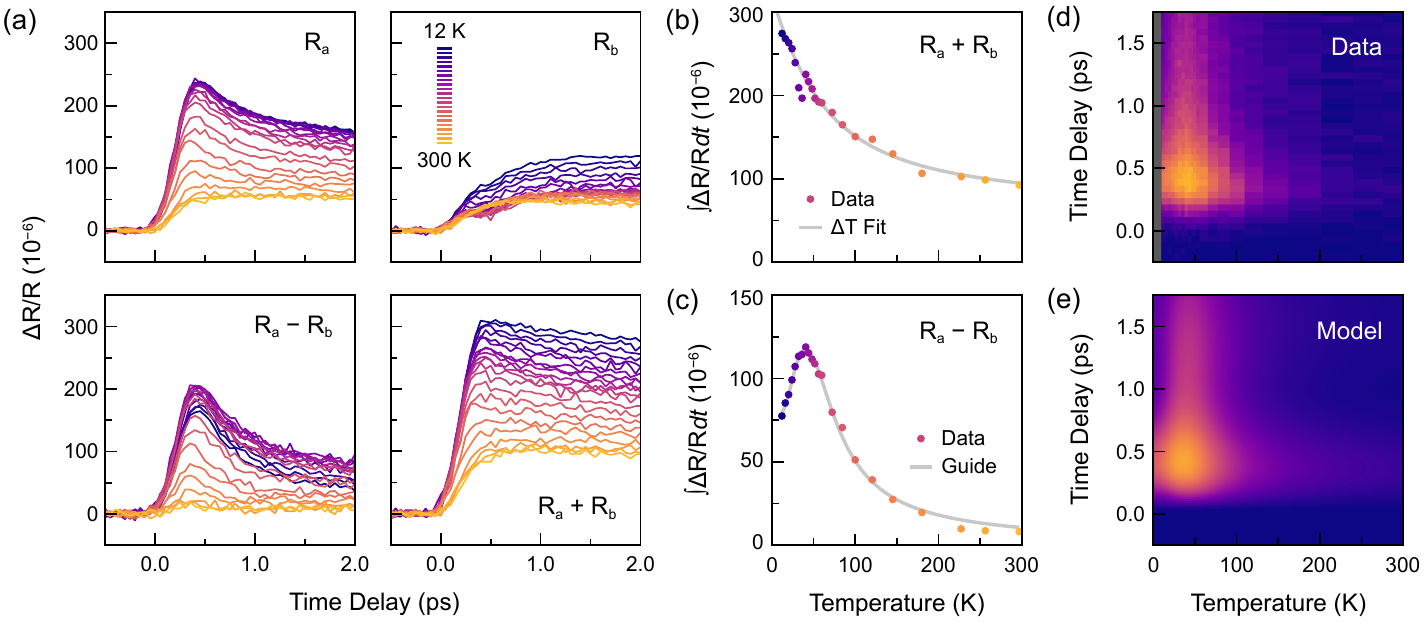}
\caption{\label{Fig2} Temperature dependence of transient reflectivity measurements. (a)~$\Delta R/R$ measurements for selected temperatures between 12~K and 300~K for polarization along the $a$-axis (upper left) and the $b$-axis (upper right), as well as their difference (lower left) and their sum (lower right). (b)~Temperature dependence of the average value of $\Delta(R_a + R_b)/R$ between 1.5 and 2~ps, representing the isotropic component of the dynamical response of the material. The response monotonically decreases with increasing temperature. The gray line is a fit to the sample temperature change, as described in the text. (c)~Temperature dependence of the average value of $\Delta(R_a - R_b)/R$ between 0 and 2~ps, representing the nematic component of the dynamical response of the material. An unusual non-monotonic response is observed. The gray line is a guide to the eye. (d)~Two-dimensional depiction of the joint time delay and temperature dependence of $\Delta(R_a - R_b)/R$. (e)~Corresponding depiction generated by fitting to the Ising model of nematic order discussed in the text.}
\end{figure*}

Now that signatures of electronic nematic order in the optical dichroism have been established, we next turn to measuring the ultrafast response of the order. As we will show, this reveals much larger signatures of nematicity than the static reflectivity and enables the extraction of dynamical information about the order. We use a standard pump-probe transient reflectivity technique to measure changes in the dichroism induced by an ultrafast ($\sim$50~fs) pump pulse of wavelength 760~nm and fluence $\sim$25~$\mu$J/cm$^2$. To eliminate any possible anisotropy due to the polarization of the pump, we keep the pump and probe pulses fixed in a cross-polarized configuration and rotate the sample about its surface normal, as illustrated in Fig.~\ref{Fig1}(c). Like the static reflectivity, we uncover a pronounced anisotropy of the transient optical response. Figure~\ref{Fig1}(d) shows raw transients of the relative change in reflectivity $\Delta R/R$ versus pump-probe time delay at 12~K for several angles between the $a$-axis (0$^\circ$) and the $b$-axis (90$^\circ$). A clear breaking of tetragonal symmetry is apparent: for optical polarization parallel to the $a$-axis, we observe a sharp peak in $\Delta R/R$ that decays over a $\sim$1~ps timescale, whereas for polarization parallel to the $b$-axis, the peak vanishes and only a flat step remains. This nematic ($C_4 \rightarrow C_2$) rotational symmetry breaking can be seen more clearly in a polar plot of the area under the transient reflectivity curve between 0 and 2~ps, as shown in Fig.~\ref{Fig1}(f). Here, no symmetrization has been performed; independent reflectivity transients were collected over a full 360$^\circ$ angular range. While the static reflectivity shows a modest 1.7\% anisotropy, the transient reflectivity exhibits a giant anisotropy exceeding 100\%. Ultrafast optical anisotropies of this magnitude are routinely observed in the nematic phase of iron-based superconductors~\cite{stojchevska2012,luo2017,liu2018,thewalt2018,lee2022}. Sub-picosecond timescales are dominated by low-energy electronic degrees of freedom, affirming that the dichroism we observe is a result of electronic nematic order rather than structural orthorhombicity.

A saturation at low temperatures is evident in the static dichroism shown in Fig.~\ref{Fig1}(b). This behavior resembles an Ising-like order parameter. To quantitatively analyze the data, we therefore develop a heuristic mean-field Ising model of the nematic order in Sr$_2$RuO$_4$. This model can describe both scenarios discussed above. Within the model, we define an order parameter $\phi$ that takes the values $+1$ or $-1$, corresponding to a nematic director oriented along the $a$- or $b$-axis, respectively. The Hamiltonian is given by
$${\mathcal{H} = -\frac{1}{2} \sum_{i,j} U(\textbf{r}_i - \textbf{r}_j) \phi_i \phi_j - F \sum_i \phi_i,}$$
where $i$ and $j$ label unit cells, $U(\textbf{r})$ is the interaction energy, related to the stiffness of the nematic order and the size of domains, and $F$ is the uniaxial strain field, which favors $+1$ domains over $-1$ domains. Within mean-field theory, the average order parameter $\left\langle\phi\right\rangle$ satisfies the self-consistency equation
$${\left\langle\phi\right\rangle = \tanh\left(\frac{U\left\langle \phi\right\rangle + F}{k_B T}\right),}$$
where $U = \sum_i U(\textbf{r}_i)/2 =  k_B T_c$ defines the critical temperature of the nematic phase transition in the absence of the external strain field $F$. Henceforth, we drop the angle brackets and call the Ising nematic order parameter $\phi$. By assuming the measured optical anisotropy is proportional to the order parameter ($R_a/R_b - 1 \propto \phi$), we can fit the data to this functional form. Such a least-squares fit, included in Fig.~\ref{Fig1}(b), yields the fitting parameters $U = 53 \pm 8$~K and $F = 53 \pm 7$~K. Reasonably good agreement with the data is achieved despite the relative simplicity of our model. Taking the model at face value, we predict a critical temperature of $T_c \approx 50$~K, where electronic nematic order would spontaneously condense following scenario~I described above. In the measured sample, however, this transition occurs gradually over a wide temperature range due to the large magnitude of the symmetry-breaking uniaxial strain field (i.e., $F \approx U$). Scenario~II, where nematic order does not occur without strain, would require $U \leq 0$, which is not consistent with our data. Based on the success of this simple Ising nematic model, we will continue to use it to extract quantitative information about our data.

Turning now to the temperature dependence of the nematicity, Fig.~\ref{Fig2}(a) shows transient reflectivity measurements at selected temperatures between 12~K and 300~K for polarization along the $a$-axis ($R_a$) and $b$-axis ($R_b$). To isolate temperature-dependent changes in the anisotropy from other effects, we also show the difference in $\Delta R/R$ between the $a$- and $b$-axis ($R_a - R_b$) as well as the sum of the two ($R_a + R_b$). From this process, we see that the sharp peak in $\Delta R/R$ is entirely due to the nematicity of the material, while the flat step after time-zero represents the isotropic component of the transient optical response. Moreover, while the isotropic response monotonically decreases with increasing temperature, as shown in Fig.~\ref{Fig2}(b), we uncover an unusual non-monotonic temperature dependence of the nematic component of the response, displayed in Fig.~\ref{Fig2}(c). The isotropic transient response is likely caused by an increase in temperature ${\Delta T}$ after absorption of the pump. Under the assumption that each pump pulse deposits a fixed amount of energy $\Delta E$ into the electronic subsystem (with heat capacity $C = \gamma T$), one can show that ${\Delta T}_\mathrm{max}(T) = \sqrt{T^2 + (2\Delta E/\gamma)} - T$. Using the value of $\gamma$ from Ref.~\citenum{mackenzie1998} and $\Delta E$ calculated from our pump fluence, we find $(2\Delta E/\gamma) \approx (80~\textrm{K})^2$ (see Supplemental Materials~\cite{SM}). A function of the form $g(T) = \alpha + \beta{\Delta T}_\mathrm{max}(T)$ with only two free parameters, $\alpha$ and $\beta$, fits the isotropic component extremely well, as shown by the gray line in Fig.~\ref{Fig2}(b). We therefore use this fact to scale the data at each temperature to match $g(T)$ and correct for laser fluctuations during the duration of the experiment, where, for example, the two data points near $\sim$35~K are otherwise anomalously low. This normalization procedure is used for the remainder of our data analysis, including Fig.~\ref{Fig2}(c).

\begin{figure}[t]
\includegraphics{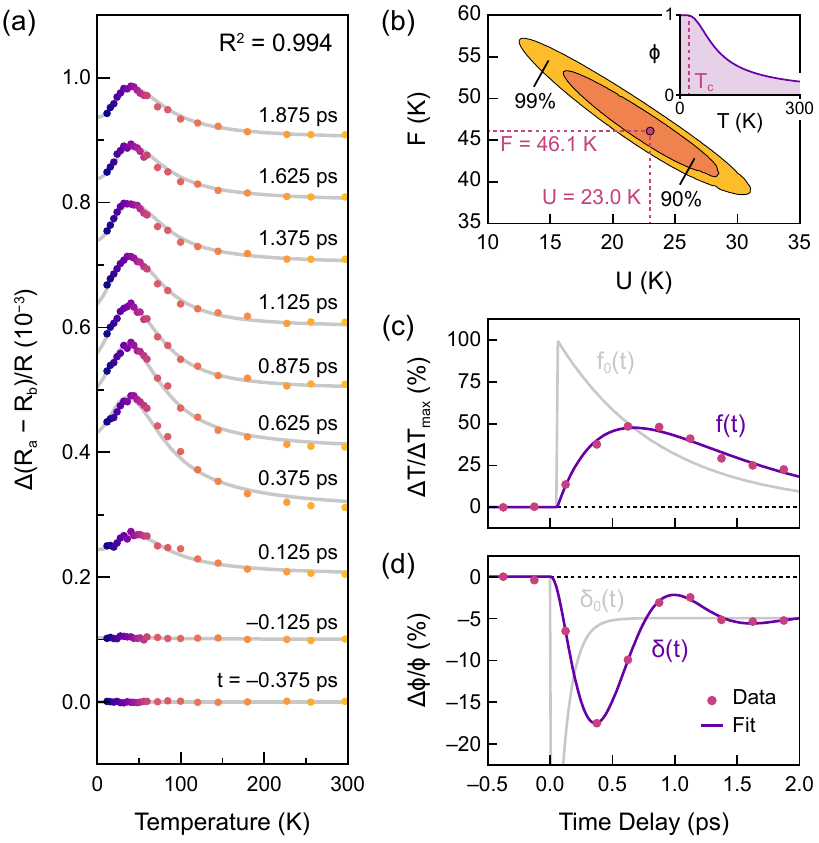}
\caption{\label{Fig3} Ising model fit of optical anisotropy. (a)~Measured $\Delta(R_a - R_b)/R$ versus temperature for selected time delays after the pump pulse. Curves are offset by $10^{-4}$ for clarity. Gray lines show the global fit to the Ising model of electronic nematic order, as described in the text. (b)~Confidence regions for the Ising parameters $U$ and $F$ computed through an $F$-test. $U > 0$ with greater than 99\% confidence, implying the existence of spontaneous nematic order. The inset shows the order parameter $\phi(T)$ together with the critical temperature $T_c$ for spontaneous nematic order when $F = 0$. (c)~Electronic temperature increase versus time delay as a fraction of the calculated maximum increase. Also shown is a fit to the solution of the heat equation $df/dt = k(f_0 - f)$ (purple curve) with background electronic temperature $f_0$ (gray curve). (d)~Relative suppression of the nematic order versus time delay, showing a coherent oscillation of the order parameter amplitude. Also shown is a fit to a damped harmonic oscillator (purple curve) with a pump-induced displacement of the equilibrium amplitude (gray curve).}
\end{figure}

Figure~\ref{Fig1}(b) shows that the onset of nematic order reduces the optical reflectivity along the $a$-axis. After the pump pulse, this reflectivity increases. It follows that the effect of the pump is to suppress the nematic order, likely through a transient increase in temperature. To model this, we assume that the dichroism is proportional to the nematic order parameter $\phi$. The measured change in dichroism is therefore proportional to the change in $\phi$:
$${\Delta(R_a - R_b)/R \propto (1 + \delta)\phi(T + \Delta T) - \phi(T),}$$
where $\phi(T)$ is the equilibrium value of the order parameter at the temperature $T$ of the measurement, $\Delta T$ is the increase in electronic temperature after absorption of the pump energy, and $\delta < 0$ represents a possible non-thermal suppression of the order parameter amplitude by the pump pulse. The previously developed Ising model provides a concrete expression for $\phi(T)$. Due to equilibration, the temperature increase $\Delta T$ is not necessarily instantaneous. Instead, we allow the temperature of the electrons participating in the nematic order to rise to some time-dependent fraction $0 \leq f(t) \leq 1$ of ${\Delta T}_\mathrm{max}$ calculated from the pump fluence as discussed above [$\Delta T(t) = f(t){\Delta T}_\mathrm{max}$]. In doing so, we assume a temperature-independent electron equilibration rate. With this model, we fit the transient optical data at all time delays and all temperatures simultaneously. The fit includes three global parameters ($U$, $F$, and a fixed scaling factor) and two time-dependent parameters [$\delta(t)$ and $f(t)$]. The results of the fit, shown in Fig.~\ref{Fig3}(a), are excellent ($R^2 = 0.994$).

A number of salient features of the data are uncovered by the fitting analysis: \textbf{(1)}~The unusual non-monotonic temperature dependence of the dichroism [Fig.~\ref{Fig2}(c)] is a result of the flattening of the nematic order parameter at low temperatures ($d\phi/dT \rightarrow 0$ as $T \rightarrow 0$), which reduces the effect of pump heating. \textbf{(2)}~Uncertainties in the fit parameters $U$ and $F$ are correlated, which is a result of their coupled influence on the functional form of $\phi(T)$. The $F$-test-derived confidence regions displayed in Fig.~\ref{Fig3}(c), however, show that $U > 0$ with high confidence. Within the applicability of our Ising model, this implies that the nematic order we observe is spontaneous (scenario~I). \textbf{(3)}~While we extract a value of $F$ that is similar to that of the static reflectivity measurement (46~K versus 53~K), the value of $U$ in the transient optical measurements is substantially smaller (23~K versus 53~K). We conjecture that repeatedly exciting the sample with high-fluence pump and probe pulses does not allow the sample to completely return to equilibrium, causing a net suppression of the order parameter. This subsequently results in a renormalization of the average interaction energy $U$. \textbf{(4)}~Figure~\ref{Fig3}(c) shows that the nematic electron temperature rises to $\sim$50\% of the maximum calculated increase ${\Delta T}_\mathrm{max}$ over a timescale of $\sim$0.5~ps. This suggests that the energy of the pump is first absorbed by background electrons during the lifetime of the pulse ($\sim$50~fs) and then is quickly transferred to the nematic electrons via thermal equilibration mediated by electron--electron interactions. At a much slower rate dictated by electron--phonon interactions, thermal equilibration with the lattice commences, causing a gradual cooling. To make this picture more quantitative, we fit $f(t)$ to the solution of the heat equation $df/dt = k(f_0 - f)$, where $f_0(t) = e^{-t/\tau}$ represents the background electronic temperature, with lattice equilibration timescale $\tau$. We find excellent agreement with the data for $\tau = 0.82$~ps. \textbf{(5)}~The non-thermal suppression of the order parameter amplitude after the pump pulse [Fig.~\ref{Fig3}(d)] reveals a coherent oscillation. This behavior can be understood within the context of a simple damped harmonic oscillator model: $d^2\delta/dt^2 + 2\gamma d\delta/dt + \omega_0^2(\delta - \delta_0) = 0$, where $\gamma$ is the damping rate, $\omega_0$ is the natural frequency of oscillation, and $\delta_0$ is the pump-induced displacement of the equilibrium order parameter amplitude. We find that a solution to this equation fits the data well if $\delta_0(t) = \alpha e^{-t/\tau} + \beta$, representing a sharp impulsive suppression at $t = 0$ followed by a rapid recovery ($\tau = 0.09$~ps) to a displaced equilibrium amplitude $\beta = -5.0$\% relative to the initial value, as shown by the gray curve in Fig.~\ref{Fig3}(d). We determine a frequency of $\omega_0/2\pi = 0.89$~THz for the amplitude mode of the nematic order.

\begin{figure}[t]
\includegraphics{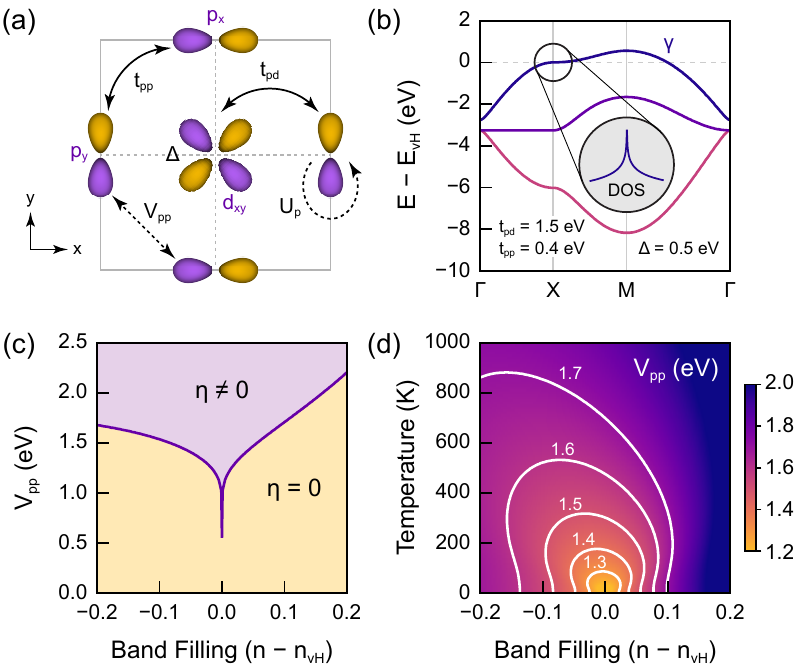}
\caption{\label{Fig4} Microscopic model of nematic order in Sr$_2$RuO$_4$. (a)~Unit cell of the RuO$_2$ plane showing the three-orbital basis (Ru $d_{xy}$ and O $p_x$ and $p_y$) of the Emery model. Hopping integrals (solid lines) and interaction parameters (dashed lines) are defined following Ref.~\citenum{fischer2011}. (b)~Resulting tight-binding band structure, with parameters taken from Ref.~\citenum{rosciszewski2015}. (c)~Interaction--filling phase diagram showing the critical value of $V_{pp}$ necessary to produce a nematic ground state ($\eta \ne 0$). (d)~Temperature--filling phase diagram for different values of $V_{pp}$. Panels (c) and (d) were calculated for $U_p = 4.4$~eV~\cite{rosciszewski2015}.}
\end{figure}

The foregoing optical experiments suggest that Sr$_2$RuO$_4$ supports a high-temperature Ising-like nematic order parameter with an order-disorder phase transition below $\sim$50~K. To put forth a possible microscopic origin for this order, we turn to the Emery model, first used to describe hole doping of CuO$_2$ planes in cuprate superconductors away from half filling~\cite{emery1987}. Here, we use the model to study intra-unit-cell nematic order in RuO$_2$ planes with fillings adjacent to the van Hove singularity of the $\gamma$~band~\cite{shen2007}. Our mean-field analysis closely follows Ref.~\citenum{fischer2011} (see Supplemental Materials~\cite{SM}). Figure~\ref{Fig4}(a) shows the hopping and interaction parameters defined within the model, and Fig.~\ref{Fig4}(b) shows the resulting band structure. Of central importance are the Coulomb repulsion between neighboring oxygen $p$-orbitals ($V_{pp}$), which favors a nematic charge order $\eta = n_x^p - n_y^p$ with different electron densities on the $p_x$ and $p_y$ orbitals, and the van Hove singularity of the $\gamma$~band at the $X$~point, which leads to a diverging density of states that supports electronic instabilities by reducing the energy cost to redistribute orbital density. Figure~\ref{Fig4}(c) shows the mean-field ground state phase diagram as a function of $V_{pp}$ and filling. A nematic phase is stabilized for sufficiently large values of $V_{pp}$ that depend on the proximity to the van Hove filling. Bulk Sr$_2$RuO$_4$ has $n - n_\mathrm{vH} \approx -0.2$~\cite{shen2007}, which requires an interaction strength $V_{pp} > 1.7$~eV to stabilize nematic order. The full temperature--filling phase diagram is displayed in Fig.~\ref{Fig4}(d), where it is seen that such order, if it exists, condenses well above room temperature. The model shows that Sr$_2$RuO$_4$ is exceptionally close to a nematic instability, and offers a plausible microscopic picture of an Ising-like ($\eta = \pm\eta_0$) nematic charge order. We conjecture that at high temperatures this nematicity is spatially disordered at the nanoscale, but shows an order-disorder phase transition to a globally nematic state at lower temperatures.

In conclusion, we have presented static and ultrafast time-resolved optical dichroism measurements of epitaxially strained thin films of Sr$_2$RuO$_4$ that strongly support the existence of electronic nematic order. By fitting our data to a simple Ising model, we conjecture that this order emerges spontaneously at low temperatures through an order-disorder transition, rather than through a large nematic susceptibility driven by strain. These results corroborate the angle-resolved transverse resistivity measurements that first uncovered electronic nematicity in Sr$_2$RuO$_4$~\cite{wu2020}. Electrical transport and optical reflectivity are disparate probes operating at different energy scales. In both cases, however, clear signatures of two-fold electronic anisotropy are evident. Furthermore, while transport can be sensitive to non-local effects such as percolation and boundary scattering, especially in a material with ostensible microscopic nematic domains, optical reflectivity is a local, bulk-sensitive probe. This difference may explain the minor discrepancies between experiments, such as the absence of nematic order in tetragonal Sr$_2$RuO$_4$/LSAT and the alignment of the nematic director with respect to the crystallographic directions. Our findings also offer a new perspective on the Fermi liquid crossover at $T_\mathrm{FL} \sim 40$~K observed by Hall transport~\cite{shirakawa1995,mackenzie1996,zingl2019}, optical spectroscopy~\cite{stricker2014}, and nuclear magnetic resonance~\cite{imai1998,chronister2022}. We speculate that the crossover is directly related to nematicity, which appears to develop at the same temperature. Indeed, such a Fermi liquid crossover has been theoretically shown to emerge in the vicinity of an Ising nematic quantum critical point~\cite{wang2019,decarvalho2019,vieira2020}. Nematic order also offers an explanation for the checkerboard charge order observed at the surface of Sr$_2$RuO$_4$ by scanning tunneling microscopy~\cite{marques2021}. Rather than surface nematic order emerging as a secondary effect, we find it likely that the checkerboard charge order is the secondary response, emerging from a combination of bulk nematicity and surface octahedral rotations. An important open question is to what extent electronic nematic order in the normal state of Sr$_2$RuO$_4$ influences the likelihood of different forms of superconductivity at lower temperatures. Future experimental and theoretical work is necessary to address this principal question.

\section*{Acknowledgments}

We thank Dr.~Ludi Miao for performing electrical resistivity measurements. This work was primarily supported by the U.S. Department of Energy, Office of Basic Energy Sciences under Award No.~DE-SC0019414. Thin film synthesis was funded in part by the Gordon and Betty Moore Foundation's EPiQS Initiative through Grant Nos.~GBMF3850 and GBMF9073 to Cornell University. Substrate preparation was, in part, facilitated by the Cornell NanoScale Facility, a member of the National Nanotechnology Coordinated Infrastructure (NNCI), which is supported by the National Science Foundation (Grant No.~NNCI-2025233). Additional support was provided by AFOSR FA9550-21-1-0168.

\end{document}